\documentclass[submitting]{nst}

\usepackage{subfigure,dcolumn}
\usepackage{epstopdf}
\usepackage{mhchem}
\usepackage{upgreek}

\begin{document}

\title{An efficient method for mapping the ${}^{12}$C+${}^{12}$C molecular resonances at low energies}

\thanks{Supported by the National Key R\&D Program of China (No. 2016YFA0400500), the National Natural Science Foundation of China (Nos. 11805291, 11475228, 11490560, 11490564, 11875329), the National Science Foundation of US (Nos. PHY-1068192, PHY-1419765), the US Department of Energy (No. DE-AC07-05ID14517), and the Fundamental Research Funds for the Central Universities (No. 18lgpy84)}

\author{Xiao-Dong Tang}
\author{Shao-Bo Ma}
\affiliation{Institute of Modern Physics, Chinese Academy of Sciences, Lanzhou, Gansu 730000, China}
\affiliation{School of Nuclear Science and Technology, University of Chinese Academy of Sciences, Beijing 100049, China}

\author{Xiao Fang}
\email[Corresponding author, ]{fangx26@mail.sysu.edu.cn}
\affiliation{Sino-French Institute of Nuclear Engineering and Technology, Sun Yat-sen University, Zhuhai, Guangdong 519082, China}

\author{Brian Bucher}
\affiliation{Idaho National Laboratory, Idaho Falls, ID 83415, USA}

\author{Adam Alongi}
\author{Craig Cahillane}
\author{Wan-Peng Tan}
\affiliation{Institute for Structure and Nuclear Astrophysics, University of Notre Dame, Notre Dame, IN 46556, USA}

\begin{abstract}
The ${}^{12}$C+${}^{12}$C fusion reaction is famous for its complication of molecular resonances, and plays an important role in both nuclear structure and astrophysics. It is extremely difficult to measure the cross sections of ${}^{12}$C+${}^{12}$C fusions at energies of astrophysical relevance due to very low reaction yields. To measure the complicated resonant structure existing in this important reaction, an efficient thick target method has been developed and applied for the first time at energies E$_{\rm c.m.}\textless$5.3 MeV. A scan of the cross sections over a relatively wide range of energies can be carried out using only a single beam energy. The result of measurement at E$_{\rm c.m.}$= 4.1 MeV is compared with other results from previous work. This method would be useful for searching potentially existing resonances of ${}^{12}$C+${}^{12}$C in the energy range 1 MeV$\textless$E$_{\rm c.m.}\textless$3 MeV.
\end{abstract}


\maketitle

\section{Introduction}

The ${}^{12}$C+${}^{12}$C is famous for its complicated molecular resonances and its importance in nuclear astrophysics~\cite{b1,b2,nsac2015}. The mechanism of these strong resonant structures have been studied and discussed by many experimental and theoretical works~\cite{b1,b2,b3,b4,b5,b6,b7,b8,b9,b10,b11,b12,b13,b14,b15}. The ${}^{12}$C+${}^{12}$C fusion reaction at low energies plays important roles in the nucleosynthesis during stellar evolution of massive stars, and is considered to ignite a carbon-oxygen white dwarf into a type Ia supernova explosion~\cite{b16,b17}. The effective energy of carbon burning is approximately from 1 to 3 MeV (Gamow window)~\cite{b16} at which the cross section varies from 10$^{-21}$ to 10$^{-7}$b. Therefore, it is extremely difficult to directly measure the ${}^{12}$C+${}^{12}$C fusion cross sections at stellar energies. Lacking a clear understanding of the complicated resonances in the ${}^{12}$C+${}^{12}$C fusion cross section, one can not reliably extrapolate the cross sections down to the unmeasured stellar energies.  Despite more than five decades of studies~\cite{b1,b2,b3,b4,b5,b6,b7,b8,b9,b10,b11,b12,b13,b14,b15}, the ${}^{12}$C+${}^{12}$C fusion cross sections at stellar energies are still highly uncertain. More precise measurements are urgently needed especially at stellar energies in order to understand the resonance-like structure and provide more reliable cross section data for the astrophysical applications. 

Both thin and thick targets have been used in experiments measuring the ${}^{12}$C+${}^{12}$C fusion cross sections. Thin carbon foils with thicknesses of a few tens of $\mu$g/cm$^2$ are usually used to measure the resonant structure and cross sections of the ${}^{12}$C+${}^{12}$C fusion reaction at relatively higher energies. However, this target suffers from carbon build-up on its surface, which increases the target thickness continuously during an experiment and brings significant discrepancies into the results. Besides that, the small cross sections at stellar energies demand a high-intensity ${}^{12}$C beam ($>10 p\mu A$). The thin carbon foils are easily damaged by such high-current beams. In the thick target approach, the beam is fully stopped inside the target, and the thick target yield is measured. The cross section is then obtained by calculating the derivative (dY/dE) from the measured thick target yield. The typical resonance width in the ${}^{12}$C+${}^{12}$C excitation function is about 50 keV or less. To precisely map the resonant structures, fine energy steps (e.g. $\Delta\textless$ 100 keV in the lab frame) are required. The yield difference between two adjacent energy points Y(E) and Y(E-$\Delta$) is calculated to determine the dY/dE. Since the thick target yield only slightly changes within a fine energy step $\Delta$, reasonably high statistics is required for each yield to get a reliable derivative for the cross section determination.

In the present article, a new thick-target approach is developed based on an analysis of the ${}^{12}$C(${}^{12}$C,$p$)${}^{23}$Na reaction. A scan of the cross sections over a relatively wide range of energies can be carried out using a single, constant beam energy. On the other hand, conventional methods require more than 10 energy points with fine steps to accomplish such a task. The new approach is much more efficient at mapping the ${}^{12}$C+${}^{12}$C resonance structure and is extremely useful in searching for new resonances at stellar energies.

The paper is organized as follows. First, we introduce a thick target experiment of ${}^{12}$C+${}^{12}$C. Second, the principle of the new thick target method is explained, and validated with a detailed Monte Carlo simulation by Geant4. Third, we apply this method to analyze the ${}^{12}$C(${}^{12}$C,$p_1$)${}^{23}$Na reaction. Fourth, the experimental results obtained with this thick target method are compared with a measurement using the traditional thin target. Finally, the strengths and weaknesses of the thick target method are also discussed.

\section{The ${}^{12}$C(${}^{12}$C,$p$)${}^{23}$N$\sc{a}$ Experiment}

The ${}^{12}$C(${}^{12}$C,$p$)${}^{23}$Na reactions were measured by experiments in the center of mass energy range of 3 MeV to 5.3 MeV using thick targets. A ${}^{12}$C beam with an intensity up to 1 p$\mu$A was provided by the 10 MV FN Tandem accelerator at the University of Notre Dame. A gas stripper system was used to enhance the intensity of the 2$^+$ charge state. The beam energies were determined by measuring the magnetic field of an analyzing magnet after the accelerator. The magnet was calibrated using the ${}^{27}$Al($p$,$n$) and ${}^{12}$C($p$,$p$) reactions.

\begin{figure*}[!htb]
\includegraphics[width=.6\textwidth]{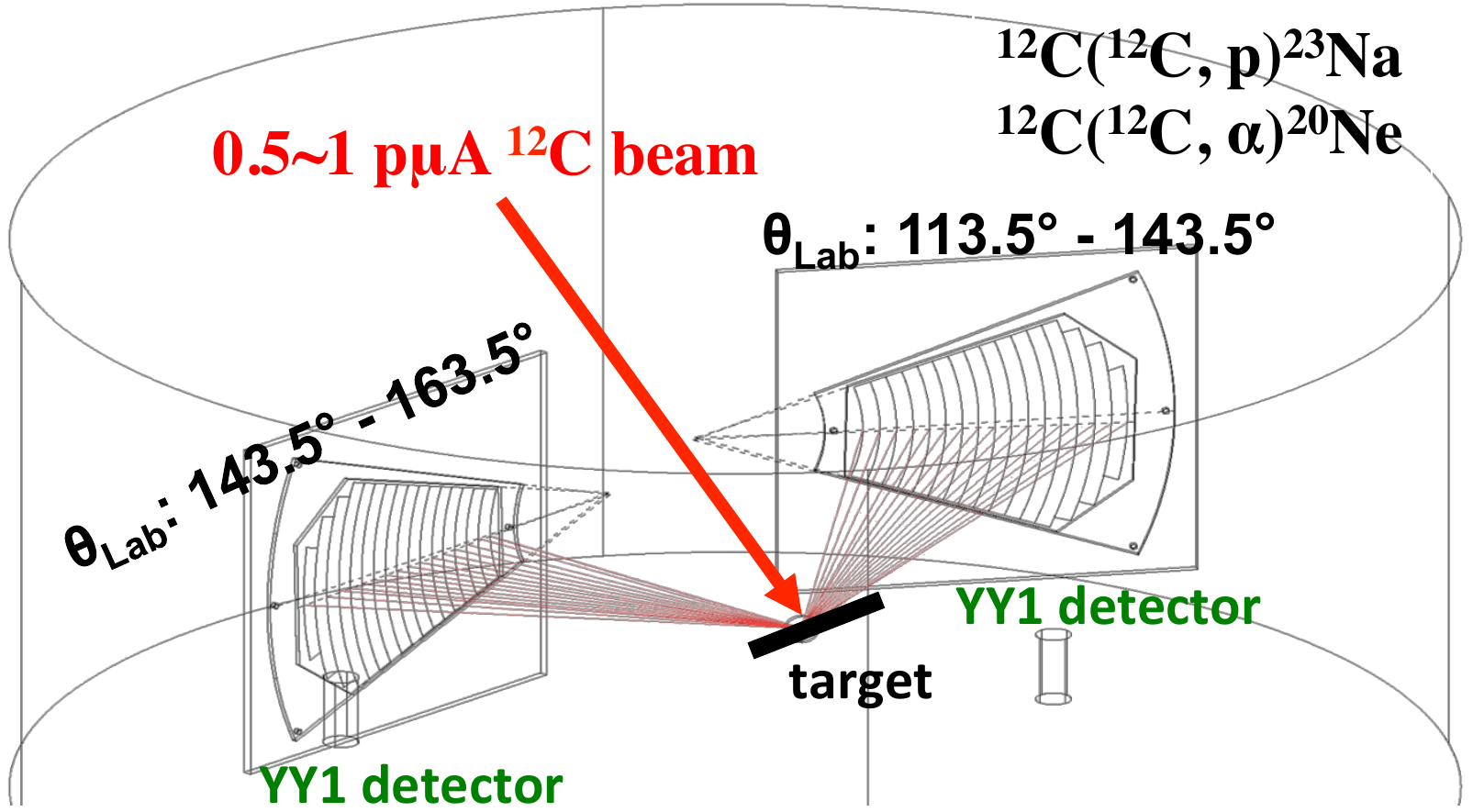}
\caption{The setup for the present experiment is shown. Two wedge shape silicon strip detectors cover the angles from 113.5$^\circ$ to 163.5$^\circ$ in the lab frame.}
\label{fig1}
\end{figure*}

The setup for the present experiment is shown in Fig.~\ref{fig1}. Two 500 $\mu$m thick YY1-type silicon detectors from Micron Semiconductor Ltd were placed at backward angles from 113.5$^\circ$ to 163.5$^\circ$ in the lab frame. For the ${}^{12}$C+${}^{12}$C fusion reaction at energies below the Coulomb barrier, the most important two reaction channels are ${}^{12}$C(${}^{12}$C,$p$)${}^{23}$Na and ${}^{12}$C(${}^{12}$C,$\alpha$)${}^{20}$Ne. Each detector was covered with a 12.7 $\mu m$-thick Al foil in the front to completely stop the $\alpha$ particles emitted from the ${}^{12}$C(${}^{12}$C,$\alpha$)${}^{20}$Ne reaction. One detector surface was perpendicular to the beam direction covering the angular range 143.5$^\circ$ to 163.5$^\circ$, and the other detector surface held a 54.4$^\circ$ angle with respect to the beam direction covering from 113.5$^\circ$ to 143.5$^\circ$. Each wedge-shaped YY1 detector was segmented into 16 strips on the front side. Thus the angular resolution for charged particles was about 1.8$^\circ$. The detectors were calibrated using a Am-Cd mixed $\alpha$ source. The energy resolution for an individual strip was about 40 keV (FWHM) for 5.486 MeV $\alpha$ particles. The total solid angle of silicon detectors was determined to be 2.59\% of 4$\pi$. A ${}^{12}$C beam with an intensity of $\sim$0.5-1 p$\mu$A was used to bombard a natural graphite target having a thickness of 1 mm.

\begin{figure}[!htb]
\resizebox{0.5\textwidth}{!}{
\includegraphics{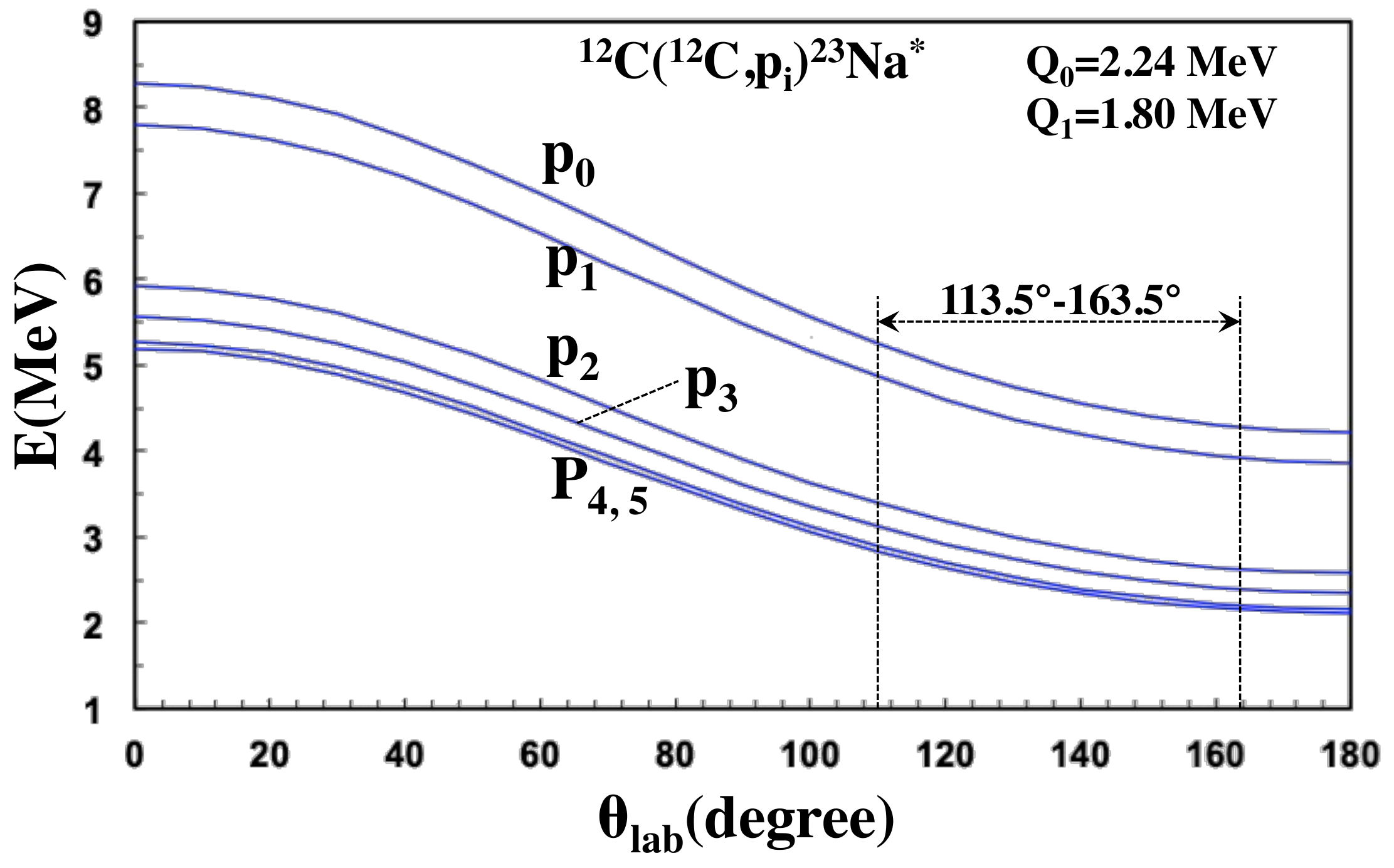}
}
\caption[width=1\textwidth]{The kinematic calculation of the ${}^{12}$C(${}^{12}$C,$p$)${}^{23}$Na reaction at E$_{lab}$=8.2 MeV. The energies of protons at different outgoing angles are given. The protons ($p_i$) associated with ${}^{23}$Na in its ground and lowest excited states are identified.}
\label{fig_kinematic}  
\end{figure}

The kinematic calculation of the ${}^{12}$C(${}^{12}$C,$p$)${}^{23}$Na reaction is shown in Fig.~\ref{fig_kinematic}. The emitted protons are labeled as $p_i$ corresponding to the $i_{th}$ excited state ($i$= 0, 1, 2, 3 $\dots$) populated in the heavy residual ${}^{23}$Na nucleus. For example, $p_0$ corresponds to ${}^{23}$Na in its ground state, and $p_1$ for the first excited state, etc. Note that $p_0$ and $p_1$ possess significantly more energy than any of the other proton groups (e.g. $p_2$,  $p_3$,  $p_4$,  $p_5$, etc.). The $\alpha$-channel has similar rules with emitted $\alpha$ particles and heavy residual ${}^{20}$Ne nuclei. 

\section{The principle of the thick target method}

Considering a reaction, 
\begin{equation}\label{eq_ab}
A + a \rightarrow B + b + Q
\end{equation}

\noindent where Q is the reaction energy which means the energy produced or absorbed by this reaction. The Q-value for reaction $A$($a$,$b$)$B$ is the total kinetic energy difference between the initial and final states. It could be determined by

\begin{equation}\label{eq_Qvalue}
Q = (\frac{M_a}{M_B}-1)E_a+(\frac{M_b}{M_B}+1)E_b-2\frac{\sqrt{M_a M_b E_a E_b}}{M_B}cos(\theta)
\end{equation}

\noindent where M$_a$,M$_b$ and M$_B$ are the masses in $amu$ of the beam, projectile, and residual particles, respectively; E$_a$ is the beam energy, and E$_b$ is the energy of the projectile particle $b$; $\theta$ is the outgoing angle of $b$. Values of E$_b$ could be measured by detectors. 

\begin{figure*}[!htb]
\includegraphics[width=.6\textwidth]{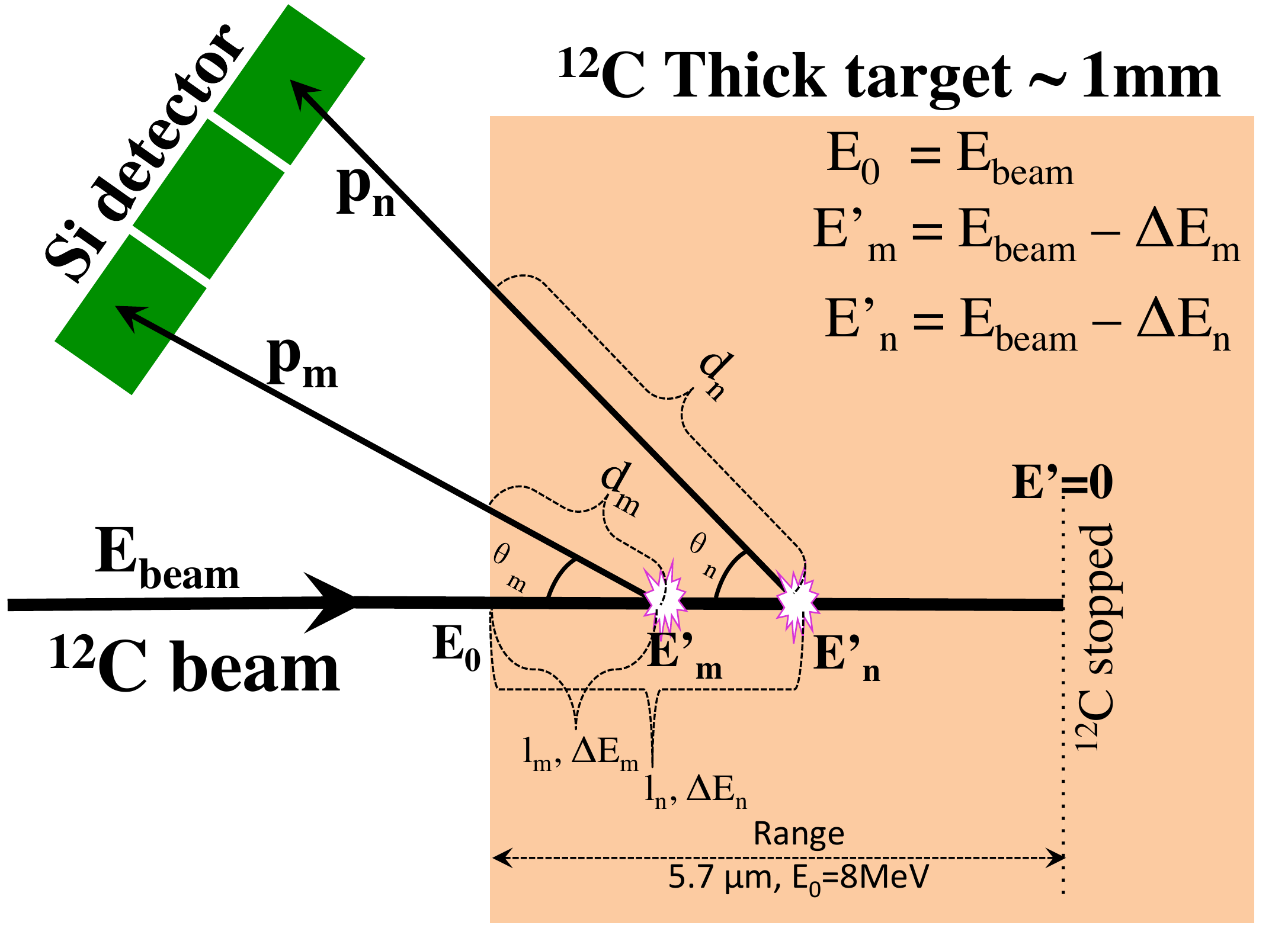}
\caption{The principle of the thick target method for ${}^{12}$C(${}^{12}$C,$p$)${}^{23}$Na reaction. For clarity, the dimensions in this figure are not drawn to scale.}
\label{fig_principle}
\end{figure*}

The principle of the thick target method for ${}^{12}$C(${}^{12}$C,$p$)${}^{23}$Na is shown in Fig.~\ref{fig_principle}. An infinitely thick target (i.e. thickness much greater than beam range inside the target material) is used in the method. The ${}^{12}$C beam particles with incident energy E$_{beam}$ bombard the target. As they collide with the target nuclei, they continuously lose energy until they either react with a target nucleus or stop within a distance of a few $\mu$ms under the front surface of the target. The range of the ${}^{12}$C beam in the ${}^{12}$C target is about 5.7 and 7.1 $\mu$m for E$_{beam}$= 8 and 10 MeV, respectively. When the ${}^{12}$C(${}^{12}$C,$p$)${}^{23}$Na reaction happens, the actual energy of the ${}^{12}$C beam is unknown. Protons produced at backward angles can easily penetrate through the target surface with an insignificant energy loss, and reach silicon strip detectors. The energies and outgoing angles of these protons are recorded by detectors. Two examples of protons, $m$ and $n$, are shown in Fig.~\ref{fig_principle}. Since the range of ${}^{12}$C beam in the ${}^{12}$C target was really small, the outgoing angles of protons (e.g. (180-$\theta_m$), (180-$\theta_n$)) were only determined by strip numbers of detectors. The Q-value is known for each individual proton group (Q= 2.24 MeV for $p_0$, Q=2.24-0.44=1.80 MeV for $p_1$, etc.), from the measured proton energies in the silicon (accounting for energy loss in the Al degrader foil) and outgoing angles (180-$\theta_m$, or 180-$\theta_n$), the actual reaction energy (E$'_m$, E$'_n$) can be reconstructed by solving Eq.~\ref{eq_Qvalue}. Therefore, a range of reaction energies [E$_{beam}$-$\Delta$E, E$_{beam}$] is scanned with a single, constant  beam energy. The effective width of the scan, $\Delta$E, usually spans from 500 to 800 keV depending on the clear identification of the reaction for events from each channel. It will be discussed later using the measurement result of ${}^{12}$C(${}^{12}$C,$p_1$)${}^{23}$Na.

The target yield derivative dY/dE is computed for each reaction energy bin after normalizing the count by the total number of incident $^{12}$C particles. The cross section for the ${}^{12}$C(${}^{12}$C,$p_i$)${}^{23}$Na reaction is then calculated from the extracted $dY/dE$ using the following equation

\begin{equation}
 \sigma (E)  =  \frac{1}{\varepsilon}\frac{{M}_{T}}{f{N}_{A}}\frac{dY}{dE}\frac{dE}{d(\rho X)}
\end{equation}

\noindent where $\varepsilon$ is the detection efficiency which is the geometric efficiency determined by an $\alpha$ source, ${M}_{T}$ is the molecular weight of the target nucleus, $f$ is the molecular fraction of the target nucleus, ${N}_{A}$ is Avogadro's number, and ${dE}/{d(\rho X)}$ is the stopping power, calculated with the SRIM code~\cite{SRIM}. 

For ease of comparison to other experimental data sets, the measured cross sections from above are converted into S$^{*}$ factors which are defined by the following equation

\begin{equation}
S^{*}=\sigma E_{c.m.} exp(\frac{87.21}{\sqrt{E_{c.m.}}}+0.46 E_{c.m.})
\end{equation}

\noindent where $\sigma$ is the cross section and E$_{c.m.}$ is the energy in the center of mass frame.

\begin{figure}[!htb]
\resizebox{0.5\textwidth}{!}{
\includegraphics{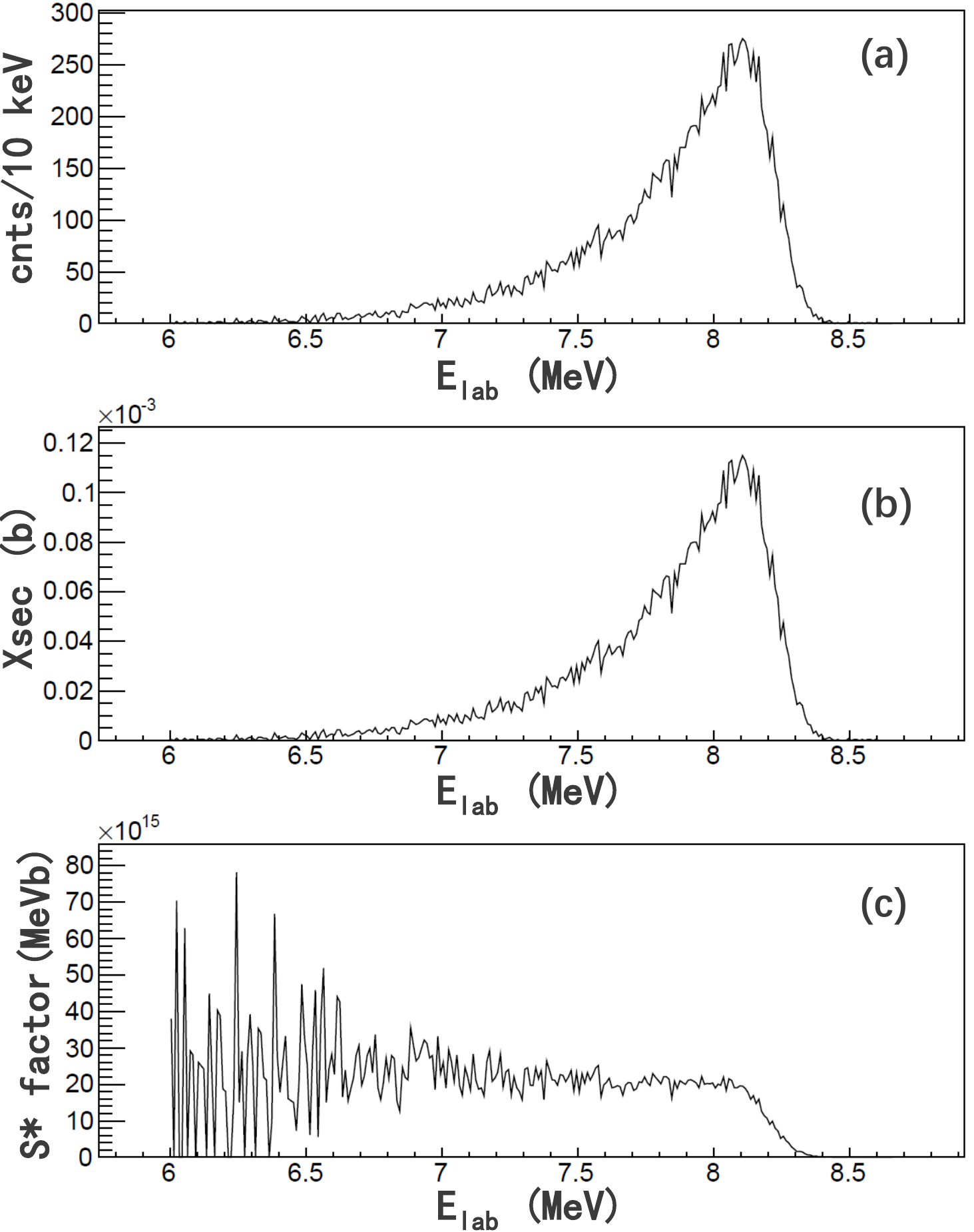}
}
\caption[width=1\textwidth]{The simulation results of a 8.2 MeV $^{12}$C beam bombarding a thick target. (a) Reconstructed reaction energy spectrum; (b) Derived cross sections; (c) Derived S$^*$ factors. }
\label{fig_simulation}  
\end{figure}

To validate the proposed thick target method, a Geant4 simulation was performed to generate a reaction energy spectrum from a constant S$^{*}$ factor input (S$^{*}$=2$\times10^{16}$ MeV*b). In the simulation, all details introduced above were considered, including geometry of the detectors, the aluminum degrader, beam straggling, etc. The reaction energy spectrum from the simulation of the $p_1$ group with an incident energy of 8.2 MeV is shown in Fig.~\ref{fig_simulation}. 

\section{Obtaining the S$^{*}$ factor of ${}^{12}$C(${}^{12}$C,$p_1$)${}^{23}$N$\sc{a}$ with the thick target method}

\begin{figure}[!htb]
\resizebox{0.5\textwidth}{!}{
\includegraphics{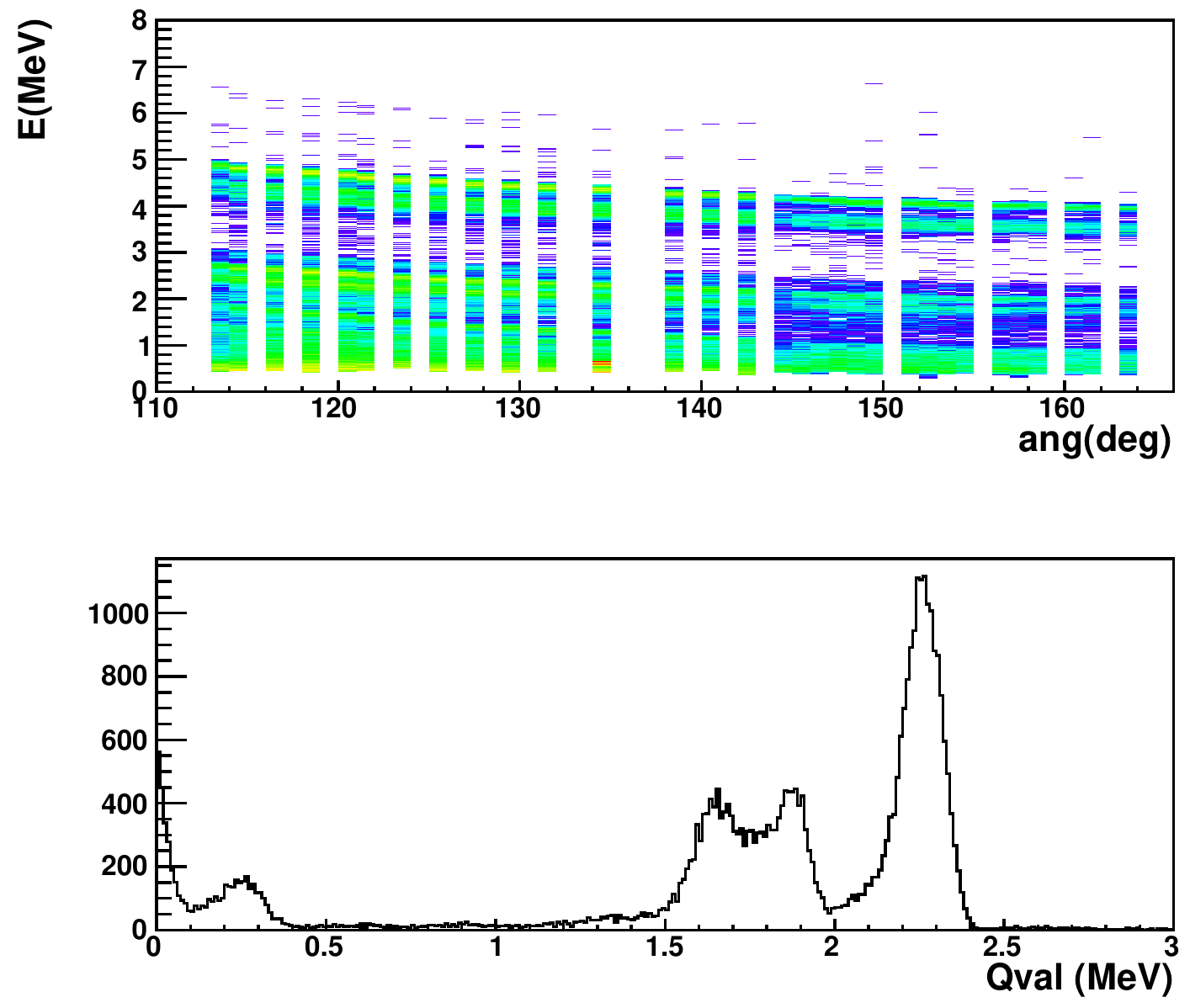}
}
\caption[width=1\textwidth]{The experimental spectra measured at E$_{beam}$= 8.2 MeV using the thick target method. (top) Energy vs. Angle of detected protons; (bottom) The Q-value for the ${}^{12}$C(${}^{12}$C,$p_{0,1,2}$)${}^{23}$Na reaction. The events between 2 and 2.4 MeV are from the $p_0$ channel. The events between 1 and 2 MeV mainly correspond to the p$_1$ channel, with a minor pollution from the $p_0$ channel.}
\label{fig:6a}      
\end{figure}

The measurement utilizing the thick target method was carried out in the energy range 3 MeV$\textless$E$_{\rm c.m.}\textless$ 5.3 MeV (6 MeV$\textless$E$_{beam}\textless$ 10.6 MeV). The measured results at E$_{beam}$= 8.2 MeV are shown in Fig. \ref{fig:6a}.

The reaction Q-value spectrum is computed from Eq.\ref{eq_Qvalue} with a constant incident energy of E$_a$= 8.2 MeV. Because the ${}^{12}$C beam loses its energy as it passes through the target medium, the actual beam energy varies from the initial incident beam energy (8.2 MeV) down to 0. As a result, the shape of the Q-value spectrum becomes much more wider and complicated than the simple sharp Gaussian shape obtained from measurement with a thin target.

The corresponding Q-values of the ${}^{12}$C(${}^{12}$C,$p_{0,1,2}$)${}^{23}$Na channels are 2.24, 1.80, and 0.164 MeV, respectively. The large Q-value difference between the $p_1$ and $p_2$ channels offers an excellent clear region for the identification of the $p_1$ events. In the present work, we focused on the analysis of the $p_1$ channel. The Q-value spectrum obtained with a thin target is expected to be narrowly 1.80 MeV. However, in the thick target method, as the reaction energy decreases in the target, the energies of the $p_1$ channel events at each fixed angle extend towards lower values as shown in the energy vs. angle plot (Fig. ~\ref{fig:6a}). The $p_1$ channel events obtained with the thick target form a wide Q-value spectrum when we compute the Q-value using a fixed beam energy, E$_a$= 8.2 MeV. 

The Q-value of the $p_0$ channel also has a lower energy tail similar to the $p_1$ channel. These low-energy events from the $p_0$ channel interfere with the high-energy events from the $p_1$ channel. As shown in Fig.~\ref{fig:6a}, this is not a big problem at E$_{c.m.}$= 4.1 MeV, because the $p_0$ cross section quickly decays with the decreasing beam energy. However, the tail of the $p_0$ channel events may contribute to the events of the $p_1$ channel around 20\% at  E$_{c.m.}$= 10.6 MeV.

\begin{figure}[!htb]
\resizebox{0.5\textwidth}{!}{
\includegraphics{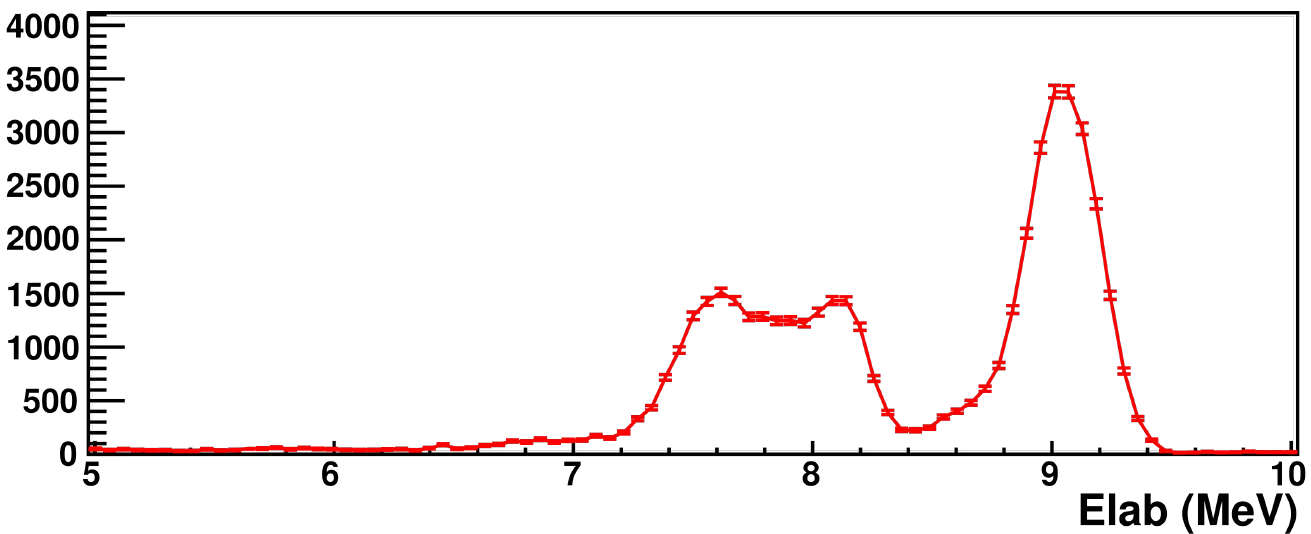}
}
\caption[width=1\textwidth]{The reaction energy spectrum constructed from the $p_1$ Q-value= 1.80 MeV. The true $p_1$ events are distributed in the range from 7 to 8.2 MeV.  All events corresponding to other channels appear with incorrect energy values.} 
\label{fig:6c}      
\end{figure}

By following the procedure introduced above, the reaction energy is computed for each event using the $p_1$ Q-value= 1.80 MeV. The reaction energy spectrum is shown in Fig.~\ref{fig:6c}. Using this unique Q-value for $p_1$, only reaction energies of events from the $p_1$ channel are correctly constructed. These events are distributed in the range of the actual reaction energies from 7.0 to 8.2 MeV. Meanwhile, all the events from other channels are placed at the $'$wrong$'$ energies. If we wanted to take a look at the $p_0$ channel, Q-value= 2.24 MeV would be used instead of 1.80 MeV to properly reconstruct the reaction energy.

The energy calibration is crucial for the determination of the actual reaction energy in this approach. Considering the uncertainties in the detector energy calibration and the degrader thickness, a minor tuning of the energy shift was applied to the reconstructed reaction energy obtained from the experiment to match the S$^{*}$ factors at the edge of the high-energy sides between the experimental and simulated spectra.

\begin{figure}[!htb]
\resizebox{0.5\textwidth}{!}{
\includegraphics{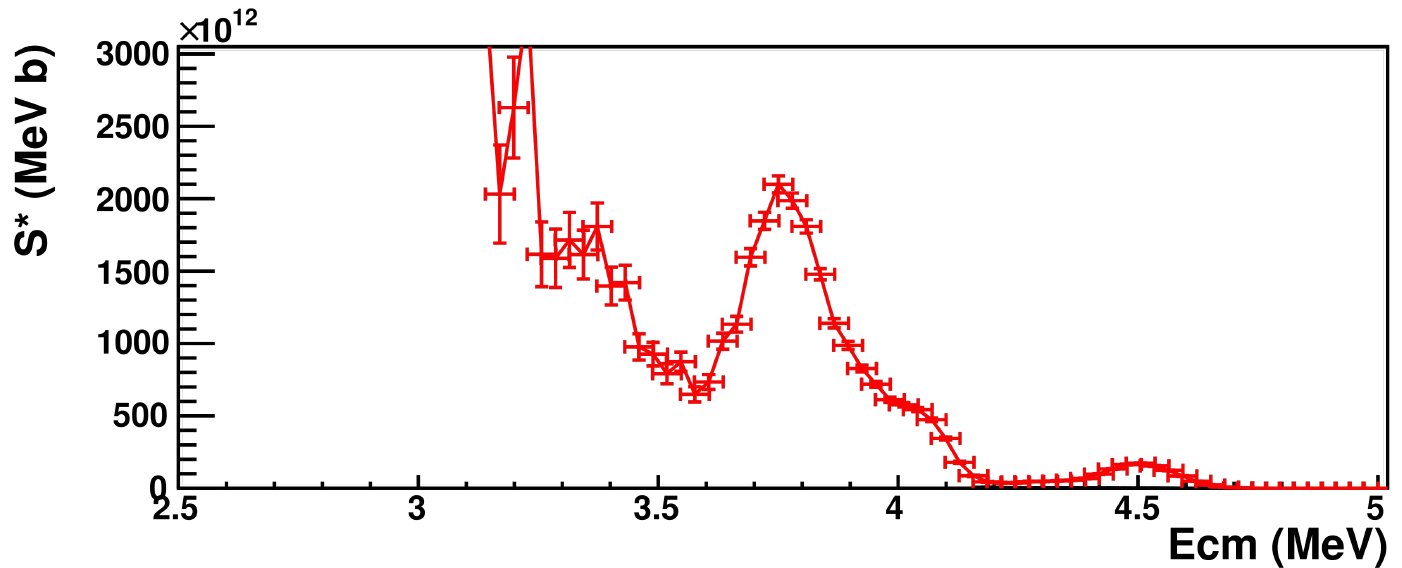}
}
\caption[width=1\textwidth]{The S$^{*}$ factor of the $p_1$ channel. The effective range of E$_{c.m.}$ is from 3.4 to 4.0 MeV.} 
\label{fig:6d}      
\end{figure}

The present S$^{*}$ factor measurement using the thick target method is shown in Fig.~\ref{fig:6d}. For the $p_1$ channel, the maximum reaction energy is E$_{c.m.}$= 4.0 MeV. The smearing of the edge at 4.1 MeV is a result of limited resolution of detectors and the spread of beam energy. The effective measurement of the S$^{*}$ factor for the $p_1$ channel stops around E$_{c.m.}$= 3.4 MeV where the background events begin to take over. These background events lead to a quickly rising S$^{*}$ factor below E$_{c.m.}$= 3.4 MeV. 

\section{Comparison with the conventional thin target measurement}

\begin{figure}[!htb]
\resizebox{0.5\textwidth}{!}{
\includegraphics{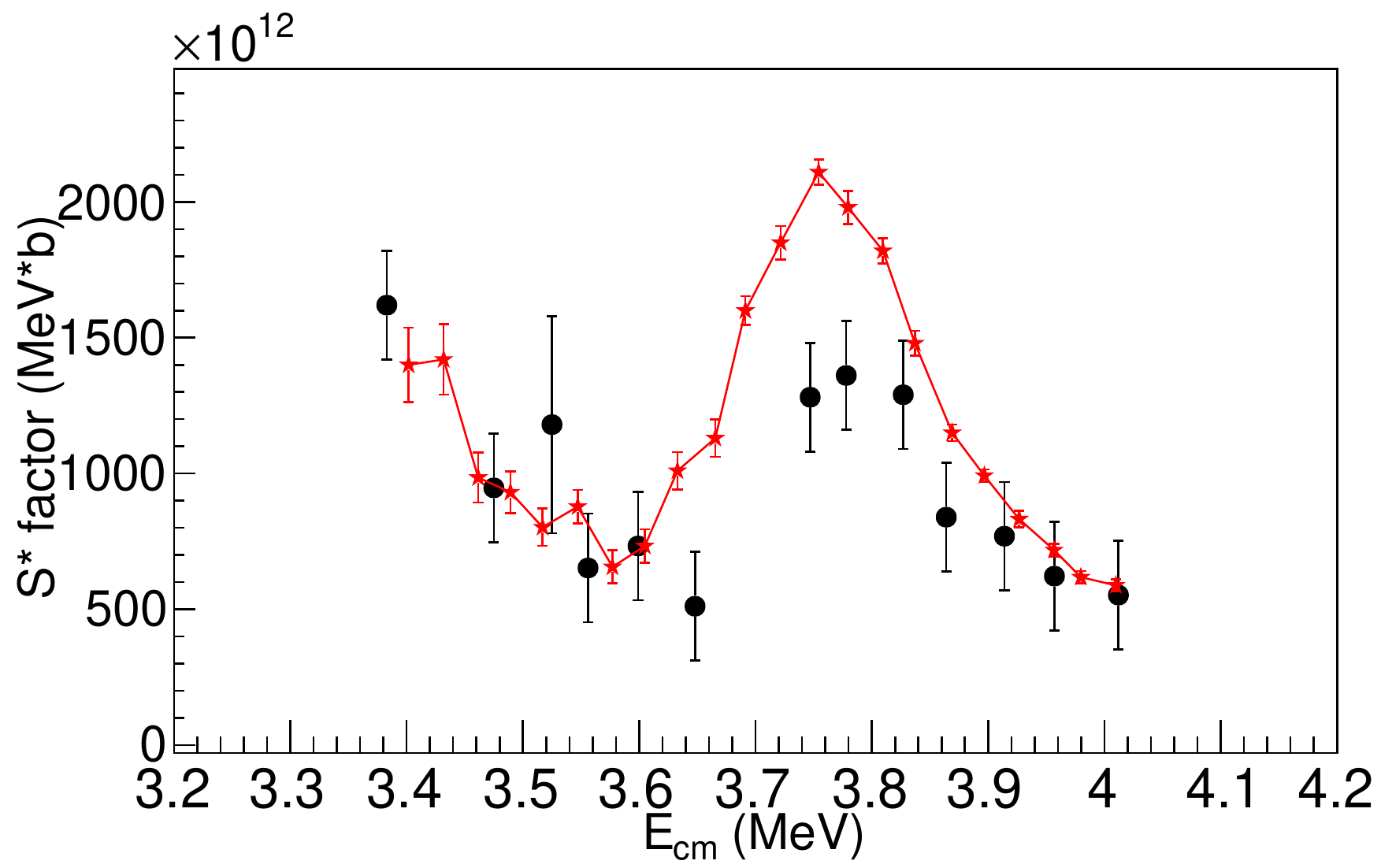}
}
\caption[width=1\textwidth]{The S$^{*}$ factor of the $p_1$ channel of ${}^{12}$C(${}^{12}$C,$p$)${}^{23}$Na. Present results are displayed in red, while the data from Ref.~\cite{b10} are displayed in black.}
\label{fig_sfactor}    
\end{figure}

The present S$^{*}$ factor of the $p_1$ channel at energies from 3.4 MeV to 4.0 MeV is shown in Fig.~\ref{fig_sfactor}, compared with an earlier measurement using the conventional thin target method by Becker $et$ $al$.~\cite{b10}. The highest reaction energy obtained by the thick target method is E$_{c.m.}$= 4.0 MeV instead of E$_{c.m.}$= 4.1 MeV because of the smearing effect resulting from the limited energy and angular resolutions. It is impressive that the complicated resonant structure of the $p_1$ channel through a wide energy range can be easily revealed with a single incident beam energy using the thick target method. A scan of about ten energy points is required if using the conventional thin target method (or with the differential thick target method). It is also noticed that the S$^{*}$ factor obtained with the thick target method is about 50\% higher than Becker's data at E$_{c.m.}$= 3.78 MeV. This is possibly an effect of the angular distribution. In the thick target experiment, we assumed a simple isotropic angular distribution because of the limited detector angular coverage. In the conventional thin target measurement of Becker $et$ $al$.~\cite{b10}, a set of silicon detectors was used to provide a well-determined angular distribution.

\section{Discussion}

The thick target method established in the present work requires a clear identification of the reaction channel of each candidate event. For the ${}^{12}$C(${}^{12}$C,$p_i$)${}^{23}$Na, both $p_0$ and $p_1$ are good candidates because of their highest proton kinetic energies and large separation from the lower Q-value channels. The other proton channels, $p_{2,3,4,5,\dots}$, are too close to each other. It is rather difficult to achieve clear identifications of each channel using only their Q-values. Therefore, the particle-gamma coincidence technique is required for identification of the Q-value~\cite{stella}. 

The ${}^{12}$C(${}^{12}$C,$\alpha$)${}^{20}$Ne reaction also could be studied by the present thick target method. The dE-E telescope technique is required to identify the $\alpha$ particles from the protons and the ambient background of $\delta$-electrons. The separation between $\alpha_0$ and $\alpha_1$, or $\alpha_1$ and $\alpha_2$ are more significant than that in the $p$ channels. This would provide a better identification of the reaction channel of each $\alpha$ event. However, the energies of these $\alpha$ particles are below 5 MeV at the backward angles, and the energy loss of $\alpha$ particles is significant inside the carbon target and Al degrader, the latter is often used in front of the detector to shield scattered ${}^{12}$C particles. A more careful correction is needed for the detected energies and angles of the $\alpha$ particles, bringing more uncertainties into the result. The associated straggling is another limitation for the $\alpha$ particle. New techniques, such as the solenoid spectrometer~\cite{b14}, are helpful for the $\alpha$-detection.

A clean background is essential for the clear identification of the reaction channel for each event. The graphite target contains the impurity D$_2$O which produces a proton background. A clean carbon target, e.g. highly ordered pyrolytic graphite (HOPG)~\cite{HOPG}, would greatly reduce contributions from target contaminants to background~\cite{b15,bucher2015,fang2017}. Direct measurements in nuclear astrophysics benefit from an underground environment, which greatly reduces cosmic ray induced background. The Jinping Underground laboratory for Nuclear Astrophysics (JUNA)~\cite{JUNA} in China, which is being constructed and expected to deliver beam in a few years, would be a proper place to do the ${}^{12}$C+${}^{12}$C measurement with the ultra-low background of the China Jinping Underground Laboratory (CJPL)~\cite{CJPL}.

The present thick target method provides an efficient way to map the resonant structure of the ${}^{12}$C+${}^{12}$C fusion reactions. An intense beam can be used with such a target to search rare events at stellar energies. A snapshot of some particular channels can be obtained efficiently with a single constant incident energy. The snapshot provides an important guidance for the following detailed energy scan using the thin target method or the differential thick target method which may reveal more details of the resonances in other reaction channels.

\section{Summary}

In summary, an efficient thick target method has been applied for the first time to measure the complicated resonant structure existing in ${}^{12}$C(${}^{12}$C,$p$)${}^{23}$Na. It can provide cross sections within a range of [E$_{beam}$-$\Delta$E, E$_{beam}$] using a single incident energy E$_{beam}$. The ${}^{12}$C+${}^{12}$C fusion reaction is one of the most important reactions in nuclear astrophysics. The efficient thick target method of the present work will be useful in searching for potentially existing resonances of ${}^{12}$C+${}^{12}$C in the energy range 1 MeV$\textless$E$_{\rm c.m.}\textless$3 MeV, where the cross sections are extremely low.

\end{document}